\documentclass[prl,twocolumn]{revtex4}

\usepackage{graphicx}

\begin{document}

\title{Quantum-Fluctuation-Driven Coherent Spin Dynamics in Small Condensates}
\author{Xiaoling Cui$^{1,2}$, Yupeng Wang$^{1}$ and  Fei Zhou$^{2}$}
\affiliation{$^{1}$Beijing National Laboratory for Condensed Matter Physics and Institute of Physics, \\
Chinese Academy of Sciences, P. O. Box 603, Beijing 100190, China \\
$^{2}$Department of Physics and Astronomy, The University of British Columbia, Vancouver, B. C., Canada V6T1Z1}
\date{{\small \today}}

\begin{abstract}
We have studied quantum spin dynamics of small condensates of cold
sodium atoms.
For a condensate initially prepared in a mean field ground state, we show
that coherent spin dynamics are {\em purely} driven by quantum fluctuations of collective spin coordinates and
can be tuned by quadratic Zeeman
coupling and magnetization.
These dynamics in small condensates can be probed in a high-finesse
optical cavity
where temporal behaviors of excitation spectra of a coupled
condensate-photon
system reveal the time evolution
of populations of atoms at different hyperfine spin states.
\end{abstract}
\maketitle

Recently, single-atom detection in optical cavities has been
realized in experiments by having atoms and cavity photons in a
strongly coupling regime\cite{Mabuchi96,Hood98}. This remarkable
achievement has been applied to study optically transported atoms in
cavities\cite{Sauer04}; furthermore the coupling between a small
Bose-Einstein condensate (BEC) and cavity photons and resultant
collective excitations have also been successfully
investigated\cite{Brennecke07}. The sensitivity that a cavity-based
atom detector has, together with a translating optical lattice which
can effectively transport ultra cold atoms from a magnetic-optical
trap to a cavity make it possible to study the physics of small
BECs. Especially, this potentially opens the door to explore
coherent dynamics of ultra-cold atoms in relatively small
condensates. The physics of BECs of small numbers of atoms can be
qualitatively different from the physics of big condensates and
represents a new domain of cold-atom research. In small condensates,
various intrinsic beyond-mean-field dynamics can be relevant within
an experimentally accessible time scale. These new physical
phenomena however have been quite difficult to study using the
standard absorption-imaging approach to cold atoms because of
relatively fewer atoms are involved in small condensates. Cavity
electrodynamics in a strong coupling regime and high sensitivities
to intra-cavity atoms on the other hand are ideal for investigating
small condensates where the beyond-mean-field dynamics are mostly
visible. In this letter, we focus on the basic concepts of
beyond-mean-field coherent spin dynamics in BECs with typically a
few tens to a few hundreds of atoms and detailed analysis of
detecting these fascinating properties of small condensates in
optical cavities with high-finesse. Research on this subject could
substantially advance our understanding of the nature of
quantum-fluctuation dynamics\cite{Song08}, in this particular case,
dynamics purely driven by fluctuations with wavelengths of the size
of condensates. Secondly, results obtained can help to better
recognize limitations of precise measurements of various interaction
constants based on mean-field coherent dynamics\cite{Chang04}.
Thirdly, our results should shed some light on the feasibility of
investigating fluctuation dynamics of small condensates using
optical cavities and also pave the way for future studies of
dynamics of coupled small condensates.

To understand spin dynamics of a small condensate,
we first study the evolution of a condensate of $N$ hyperfine spin-one sodium atoms which is
initially prepared in a mean field ground state,

\begin{eqnarray}
|{\bf n}\rangle=\frac{({\bf n}\cdot
\psi^{\dag})^N}{\sqrt{N!}}|0\rangle. \label{initial}
\end{eqnarray}
Here ${\bf n}$ is a unit director and three components of
$\psi^\dag$, $\psi^\dag_\alpha$, $\alpha=x,y,z$ are creation
operators for three spin-one states,
$|x\rangle=(|1\rangle-|-1\rangle)/\sqrt{2}$,
$|y\rangle=(|1\rangle+|-1\rangle)/i\sqrt{2}$ and
$|z\rangle=|0\rangle$ respectively. And in this representation,
$S_\alpha=-i\epsilon_{\alpha\beta\gamma} \psi^\dag_\beta\psi_\gamma$
is the total spin operator. States in Eq.\ref{initial} with ${\bf
n}={\bf e}_z$ minimize the interaction energy of the following
Hamiltonian for spin-one atoms in the presence of a
quadratic Zeeman coupling along the $z$-direction,

\begin{equation}
H=\frac{c_2}{N}\mathbf{S}^2+q(\psi^\dag_x\psi_x+\psi^\dag_y\psi_y).
\label{Hamiltonian}
\end{equation}
Here $c_2$ is a spin interaction constant and $q$ is the quadratic
Zeeman coupling\cite{Ho98,Ohmi98,Law98,Stenger98}. Mean field ground
states are stationary solutions to the multi-component
Gross-Pitaevskii equations for spin-one atoms and dynamics of these
initial states demonstrated below are therefore a beyond-mean field
phenomenon. When deriving Eq.\ref{Hamiltonian} for a trapped
condensate, we assume that spin dynamics are described by a single
mode, i.e. $\psi_\alpha({\bf r},t)=\sqrt{\rho({\bf
r})}\psi_\alpha(t)$; for a small condensate of less than one
thousand weakly interacting atoms, this approximation is always
valid. $c_2$ is typically a few nano kelvin for sodium atoms;
$q=(\mu_B B)^2/(4\Delta_{hf})$ and the hyperfine splitting is
$\Delta_{hf}=(2\pi)1.77 GHz$ ($\mu_B$ is the Bohr magneton and
$\hbar$ is set to be unity).

To illustrate the nature of non-mean-field dynamics and crucial role
played by quantum fluctuations,
we expand the full Hamiltonian in Eq.\ref{Hamiltonian}
around a mean field ground state.
In the lowest order expansion, we approximate $\psi^\dag
\approx \sqrt{N} {\bf e_z}+\psi^\dag_{x}{\bf
e}_x+\psi^\dag_y {\bf e}_y$, and $\psi^\dag_{x,y}$ are much less than
$\sqrt{N}$; the Hamiltonian then can be expressed
in terms of the bilinear terms

\begin{equation}
H_B=\sum_{\alpha=x,y} \frac{q+4 c_2}{2N} P_\alpha^2+\frac{q N}{2}
\theta_\alpha^2+...
\label{HO}
\end{equation}
where for $\alpha=x,y$,
$\theta_\alpha =\frac{1}{\sqrt{2N}} (\psi^{\dag}_\alpha+
\psi_\alpha)$ and
$P_\alpha =i\sqrt{\frac{N}{2}}(\psi^{\dag}_\alpha-
\psi_\alpha)$
are pairs of conjugate operators which satisfy the usual commutation
relations
$[\theta_\alpha, P_\beta]=i\delta_{\alpha,\beta}$.
Semiclassically, collective coordinates
$\theta_\alpha$, $\alpha=x,y$
represent projections of $\psi^{\dag}$ or order parameter ${\bf n}$  in
the $xy$ plane, and $P_{x(y)}\sim S_{y(x)}$ is the spin projection along the $y(x)$-direction.
The bilinear Hamiltonian is
equivalent to a harmonic
oscillator moving along the direction of $\theta_{x,y}$ with
a mass $m_{eff}=\frac{N}{q+4 c_2}$, a harmonic
oscillator frequency $\omega=\sqrt{q(q+4 c_2)}$ and effective spring constant
$q N$;
the mass at $q=0$ is induced by scattering between atoms.
The excitation spectrum is $E_n=(n+1/2)\omega$, $n=0, 1,2...$ .
When $q=0$, the Hamiltonian describes a particle moving in a free space.

\begin{figure}[ht]
\includegraphics[height=9cm,width=\columnwidth]{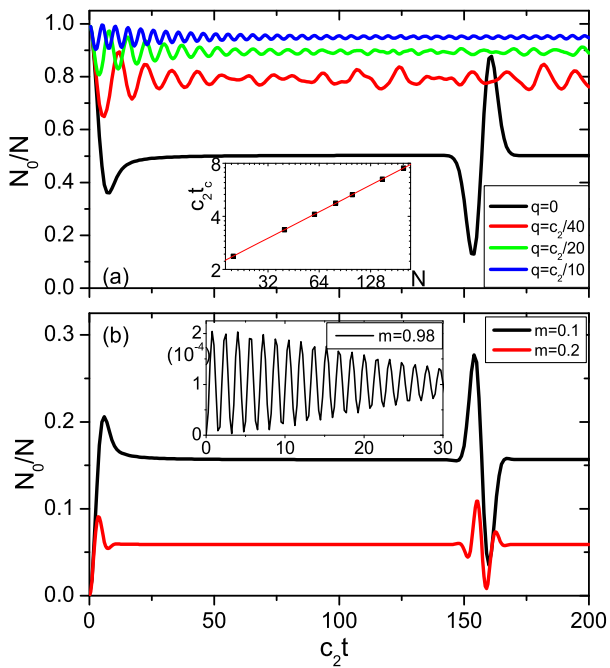}
\caption{a) (color online) Time evolution of $N_0(t)$, the atom population at
$|1,0\rangle$ state for different quadratic Zeeman coupling $q$.
Initially, all $N=200$ atoms occupy $|1,0\rangle$
state which corresponds to a mean field ground state. The inset is
$t_c$, the time for the first dip in the $q=0$ data, as a function
of the number of atoms $N$. b) Time evolution of $N_0(t)$ for
different magnetization $m$ (here $q=0$). All initial states are
again mean-field ground states for given $m$. Inset is for $m=0.98$.
In this and Fig.\ref{fig4},\ref{fig5}, $c_2=(2\pi) 50Hz$. }
\label{fig1}
\end{figure}

In the ground state of the bilinear Hamiltonian of Eq.\ref{HO},
$\langle\theta_\alpha\rangle$ $=\langle P_\alpha\rangle=0$ and ${\bf
n}$ and $\langle{\bf S}\rangle$ have no projections in the $xy$
plane. However, quantum fluctuations of $\theta_{x,y}$-coordinates
in the ground state can be estimated as
$\langle\theta_{\alpha}\theta_\alpha\rangle$ $=\frac{1}{2N}
\sqrt{\frac{q+4c_2}{q}}$. This is a measure of how strongly ${\bf
n}$ fluctuates in the $xy$-plane. As expected, these quantum
fluctuations are substantial only when $q$ is small and are
suppressed by a quadratic Zeeman field which effectively pins the
order parameter along the $z$-direction. A direct calculation also
shows that the amplitude of quantum fluctuations
$\langle\theta^2_\alpha\rangle_{MF}$ in the {\em mean field ground
state} defined in Eq.\ref{initial} is $1/2N$. This indicates that
the mean field ground state is a good approximation only when $q \gg
4 c_2$. On the other hand, as $q$ decreases and the effective spring
constant gets smaller, the deviation becomes more and more severe.
When $q$ approaches zero, quantum fluctuations $\theta_\alpha$ in
the harmonic oscillator ground state become divergent implying that
the mean field ground state is no longer a good approximation.

Indeed, the the energy of mean field ground state is
$E_{MF}=\frac{q}{2}+c_2$ which is much higher than
$\frac{1}{2}\omega$ when $q \ll c_2$; such a state corresponds to a
highly excited wave packet, because of a relatively narrow spread
along $\theta_\alpha$-directions and consequently an enormous
kinetic energy associated with momenta $P_\alpha$. We therefore
expect that dynamics in this limit could dramatically differ from a
stationary solution. Since the total number of atoms $N$ is equal to
$\sum_{\alpha} \psi^{\dag}_\alpha\psi_\alpha$, the population of
atoms at $|z\rangle$ (or $|1,0\rangle$) state
$N_0=\langle\psi^\dag_z\psi_z\rangle$ is directly related to quantum
fluctuations of $\theta_\alpha$ and $P_\alpha$,

\begin{eqnarray}
N_0=N+{1} -\sum_\alpha \left( \frac{N}{2}
\langle\theta_\alpha^2\rangle+\frac{1}{2N}\langle
P_\alpha^2\rangle\right). \label{fluc}
\end{eqnarray}
Eq.\ref{fluc} shows that the time evolution of $N_0(t)$ is
effectively driven
by quantum fluctuations in $\theta_\alpha$ and
$P_\alpha$; a study of $N_0(t)$ probes underlying quantum-
fluctuation dynamics.

For an initial state prepared in a mean field ground state with
${\bf n}={\bf e}_z$ where all atoms condense in $|1,0\rangle$ state,
one finds that $\langle \theta^2_\alpha\rangle= \frac{1}{2N}$ and
$\langle {P^2_\alpha}\rangle=\frac{N}{2}$. The evolution of such a
symmetric Gaussian wave packet subject to the bilinear Hamiltonian
can be solved exactly using the standard theory for harmonic
oscillators. The wave packet will remain to be a Gaussian one with
the width oscillating as a function of time. Qualitatively, because
of the symmetry, only harmonic states with even-parity are involved
in dynamics and therefore the oscillation frequency is $2\omega$.
Furthermore during oscillations, the kinetic energy stored in
initial wave packets is converted into the potential one and {\em
vice versa}. Especially when $q \ll c_2$, oscillations are driven by
the enormous initial kinetic energy associated with $P_\alpha$; the
oscillation amplitude can be estimated by equaling the total energy
$E_{MF}$ to the potential energy which leads to
$\langle\theta_\alpha^2\rangle$ $\sim c_2/(Nq)$. A straightforward
calculation yields the time dependence of
$\langle\theta^2_\alpha\rangle$ and $\langle P^2_\alpha\rangle$ that
leads to

\begin{equation}
\frac{N_0}{N}=1-\frac{8 c_2^2}{q(q+4c_2)N}\sin^2wt.
\label{oscillation}
\end{equation}
The oscillating term in Eq.\ref{oscillation} shows the deviation
from the stationary behavior due to quantum fluctuations in
$\theta_\alpha$-coordinates. The deviation is insignificant when $q$
is not too small; however when $q$ is of the order of $c_2/N$, we
expect that the non-mean field dynamics becomes very visible. Note
that the approach outlined here neglects all higher order anharmonic
interactions and therefore is only valid when the relative amplitude
of fluctuations is small; that is when $q \gg c_2/N$.

When $q$ approaches zero, the short time dynamics following the
bilinear Hamiltonian is equivalent to a particle of a mass
$m_{eff}=N/4c_2$ that is initially localized within a spread
$\langle\theta_\alpha^2\rangle=1/2 N$ having a ballistic expansion
with a typical velocity give as ${\langle
v^2_\alpha\rangle}=8c^2_2/N$. The time dependence of spread
$\langle\theta^2_\alpha\rangle$ therefore is $1/2N + (8c^2_2/N)
t^2$. So at $t \sim \sqrt{N}/c_2$, the number of atoms not occupying
the initially prepared $|1,0\rangle$ state becomes of order of $N$.
This limit was first addressed by Law {\em et al.} in the context of
four-wave-mixing theory\cite{Law98}, and also in early
works\cite{Zhou01,Diener06}; to describe the physics after this
characteristic time scale requires analysis of full quantum
dynamics. This time scale however becomes quite long for a few
million atoms which makes it difficult to observe quantum dynamics
in large condensates.

In the following, we are going to present our numerical results on
dynamics and focus on its dependence on quadratic Zeeman coupling
$q$ and magnetization $m$. For a condensate of $N=200$ atoms, we
numerically integrate the time-dependent N-body Schrodinger equation
of the quantum Hamiltonian in Eq.\ref{Hamiltonian}. The time
evolution of $N_0$ driven by quantum fluctuations is shown in
Fig.\ref{fig1}a). As $q$ increases far beyond $0.2 c_2$, $N_0$
oscillates as a function of time with frequency $2\omega$ and the
amplitude of oscillations decreases; the damping is not visible over
tens of oscillations. When $q$ is below $0.2 c_2$, anharmonic
effects become substantial and oscillations are no longer perfect;
when $q=c_2/40$, oscillations are strongly damped after a few cycles
and revived afterwards. For $q=0$, $N_0$ drops to a minimum of about
$0.38 N$ when $t=t_c=0.53 \sqrt{N}/c_2$ and remains to be a constant
before reviving to be $0.8 N$ at about $10 t_c$. For sodium atoms
with a typically density $2\times 10^{14} cm^{-3}$,
$c_2=(2\pi)50 Hz$; $t_c=23.8ms$ for $N=200$ and
increases to a few seconds when $N$ reaches $2 \times 10^6$.

We have also studied the quantum dynamics of a mean
field condensate with a finite magnetization along the
$z$-direction, ${\bf m}=m {\bf e}_z$. States which minimize
the mean field energy of the Hamiltonian in Eq.\ref{Hamiltonian} with $q=0$ are

\begin{eqnarray}
|m\rangle=\frac{[(\cos\eta {\bf e}_x +i\sin\eta {\bf e}_y)\cdot
\psi^\dag]^N} {\sqrt{N!}}|0\rangle \label{mstate}
\end{eqnarray}
where $\sin 2\eta=m$, $m (\in [-1,1])$
is the normalized magnetization.
By expanding the Hamiltonian around these mean field states, one
obtains a harmonic oscillator Hamiltonian
defined in terms of conjugate operators,
$\theta_z=\frac{1}{\sqrt{2N}}(\psi^{\dag}_z+\psi_z)$ and
$P_z=i\sqrt{\frac{N}{2}}(\psi^{\dag}_z -\psi_z)$. The effective mass
is $m_{eff}=\frac{N}{2c_2 (1+\sqrt{1-m^2})}$ and the harmonic
oscillator frequency $\omega=2 |m| c_2$. States shown in
Eq.\ref{mstate} have a narrow width along the direction of
$\theta_z$, $\langle\theta^2_z\rangle=1/2N$ and therefore carry
large conjugate momenta $P_z$; the corresponding large kinetic
energy drives a unique non-mean field quantum spin dynamics. The
harmonic expansion again is only valid when $m$ is large and
fluctuations are weak. Simulations of the full Hamiltonian have been
carried out in this case; in Fig.\ref{fig1}b), we show the time
dependence of $N_0(t)$ for different magnetization. Only when $m$ is
close to unity, weakly damped oscillations are observed.

\begin{figure}[ht]
\includegraphics[width=\columnwidth]{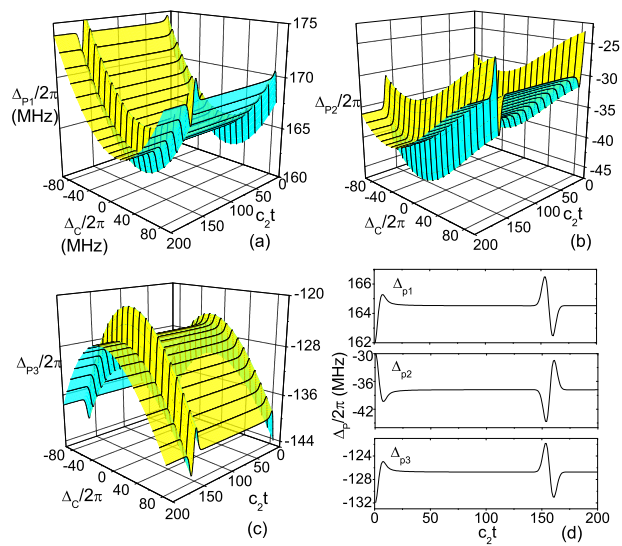}
\caption{(a,b,c) (color online) Eigenfrequencies $\Delta_{p}$ as a function of
$t$ for different detuning $\Delta_c$ when the relative
population at state $|1,0\rangle$, $\rho_0 (t)$ evolves. At $t=0$,
all atoms occupy $|1,0\rangle$ state and $N=200$. d) $\Delta_p$ as a
function of time $t$ for $\Delta_c=0$.} \label{fig4}
\end{figure}

We propose a method to probe quantum spin dynamics
of a small condensate of spin-one sodium atoms using cavity quantum
electrodynamics.
For a Bose-Einstein condensate with N
atoms coupled to a quantized field of a cavity, a single cavity photon can
coherently interact with
atoms which leads to a collective coupling of $g\sqrt{N}$\cite{Tavis68}.
In experiments\cite{Sauer04,Brennecke07},
atoms are transported into a cavity via a moving optical
lattice; excitations are measured by individual
recordings of cavity transmission when frequencies of
an external probe light are scanned.
Here we consider a multi-component BEC coupled to a single cavity mode;
the eigenfrequencies of the
coupled system uniquely depend on populations at three hyperfine
states.
By measuring the energy spectrum of this coupled system, one obtains
temporal behaviors of
atom populations at different states.

We restrict ourselves to excitations
which involve a single cavity photon interacting with atoms in a BEC.
We study atomic transitions from $3S_{\frac{1}{2}}\rightarrow
3P_{\frac{1}{2}}$ in sodium atoms. The Hamiltonian
consists the following terms,
\begin{eqnarray}
{H}_{cavity}&=&\sum_i \hbar w_{g_i}\hat{g}_i^{\dag}\hat{g}_i+\sum_j \hbar
w_{e_j}\hat{e}_j^{\dag}\hat{e}_j+\sum_p \hbar w_c
\hat{c}_p^{\dag}\hat{c}_p\nonumber\\
&&-i\hbar \sum_p\sum_{i,j} g_{ij}^p
\hat{e}_j^{\dag}\hat{c}_p\hat{g}_i+h.c.,
\end{eqnarray}
where $i$ labels three states $|F=1, m_{F}=0,\pm 1\rangle$ in
$3S_{\frac{1}{2}}$ orbital and $j$ eight states $|F'=1,
m_{F'}\rangle$, $|F'=2, m_{F'}\rangle$ in $3P_{\frac{1}{2}}$
orbital. $\hat{g}_i^{\dag}$ and $\hat{e}_j^{\dag}$ create atoms in
one of $3S_{\frac{1}{2}}$ and $3P_{\frac{1}{2}}$ states respectively
with corresponding frequencies $w_{g_i}$,$w_{e_j}$.
$\hat{c}_p^{\dag}$ creates a photon with frequency $w_c$ and
polarization $p$ in the cavity mode. $g_{ij}^p(=D_{ij}^p\sqrt{\hbar
w_c/2\epsilon_0V})$ is the coupling strength for a transition
$i\rightarrow j$ driven by a cavity photon with polarization $p$,
which depends on the dipole matrix element $D_{ij}^p$, the effective mode
volume $V$.

\begin{figure}[ht]
\includegraphics[width=\columnwidth]{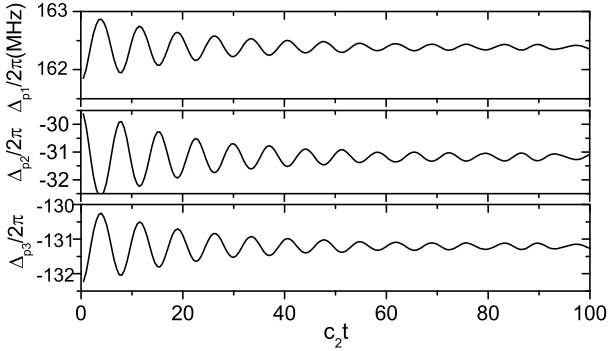}
\caption{Eigenfrequencies $\Delta_p$ as a function of time $t$
driven by dynamics of population $\rho_0(t)$ in the presence of
quadratic Zeeman coupling $q=0.05 c_2$ (or $B=95 mG$). Again at
$t=0$, all atoms occupy $|1,0\rangle$ state and $N=200$,
$\Delta_c=0$.} \label{fig5}
\end{figure}

For simplicity, we set the energy of $3S_{\frac{1}{2}}$ states to be
zero, i.e. $w_{g_i}=0$; the energy of excited states is
$w_{e}^{1,2}=w_a\pm \Delta$ with $2\Delta$ being the hyperfine
splitting between $F'=2$ and $F'=1$ states. For atomic transitions
induced by left-circularly($\sigma^+$) polarized photons, the
selection rule is $\Delta F=0, \pm 1, \Delta m_F=1$. In a cavity, a
state with a cavity photon (labeled as 1c),$N_{m_F}$ atoms at
$3S_{\frac{1}{2}}$ $|1, m_F\rangle$ states and no atoms at excited
states (labeled as $0_j$) is expressed as $|1_c; N_1,N_0,N_{-1};
0_j\rangle$; it is coupled to the following states with one of atoms
excited to $3P_{\frac{1}{2}}$ states (labeled $1_{F'}$) and no
cavity photons (as $0_c$), $|0_c; N_1-1,N_0,N_{-1};
1_{F'=2}\rangle$, $|0_c; N_1,N_0-1,N_{-1}; 1_{F'=1}\rangle$, $|0_c;
N_1,N_0-1,N_{-1}; 1_{F'=2}\rangle$, $|0_c; N_1,N_0,N_{-1}-1;
1_{F'=1}\rangle$, $|0_c; N_1,N_0,N_{-1}-1; 1_{F'=2}\rangle$. We
diagonalize the Hamiltonian matrix and obtain six eigenfrequencies
$\omega_p$ for this coupled system. Three are $3P_{\frac{1}{2}}$
orbitals without mixing with $3S_{\frac{1}{2}}$ states, $w_p=w_a\pm
\Delta$; the other three depend on {\em relative} populations of
atoms at each spin state, $\rho_{m_F}=N_{m_F}/N$, $m_F=0,\pm 1$. The
latter three eigen frequencies are determined by the eigen value
equation,
\begin{eqnarray}
(\Delta_p-\Delta_c)(\Delta_p^2-\Delta^2)-\Delta_p
Ng_1^2 F_1-\Delta
Ng_1^2 F_2=0.\label{eigen}
\end{eqnarray}
Here $\Delta_p=w_p-w_a$, $m=\rho_{+1}-\rho_{-1}$ is the normalized
magnetization, and $g_1$ is the coupling between $3S_{\frac{1}{2}}$
$|F=1, m_{F}=1\rangle$ and $3P_{\frac{1}{2}}$ $|F'=2,
m_{F'}=2\rangle$ by $\sigma^+$ light; $F_1=(2+m)/3$ and
$F_2=(1+m)/2-\rho_0/6$. Apparently eigenfrequencies
$\Delta_{p1,p2,p3}$ are a function of $\rho_{m_F}$ and therefore can
be used to probe the variation in $\rho_{0,\pm 1}$ due to coherent
spin dynamics. $\Delta_{p1,p2,p3}$ depend on a
dimensionless parameter
$r=\frac{\sqrt{6}}{3}\frac{\sqrt{N}g_1}{\Delta}$. When detuning
$\Delta_c=0$ and as $r\rightarrow \infty$, $\Delta_{p1,p2,p3}$ are
around $0, \pm \frac{\sqrt{6}}{3}\sqrt{N}g_1$ respectively.
Eigenfrequency $\Delta_{p2}$ varies from $-3\Delta/4$ to $-\Delta/2$
when $\rho_0$ increases from $0$ to $1$; the variation amplitude
$\delta=\Delta_{p}(\rho_0=1)-\Delta_{p}(\rho_0=0)$ reaches a
saturated value $\Delta/4$. For sodium atoms,
$\Delta=(2\pi)94.4MHz$; cavity parameters are chosen according to
Ref.\cite{Brennecke07} and $g_1=(2\pi)10MHz$. In  Fig.
(\ref{fig4}), we show the evolution of $\Delta_p$ in time for
different detuning $\Delta_c$ when atoms are initially prepared at
state $|1,0\rangle$ of $3S_{1/2}$. The evolution of $\Delta_p(t)$
which can be probed by a $\sigma^+$ beam
maps out population $N_0(t)$
driven by underlying quantum fluctuations. In Fig. (\ref{fig5}), we
further show the time dependence of $\Delta_p$ due to oscillatory
quantum spin dynamics for $q=0.05c_2$($B \approx 95mG)$, $N=200$ and
$\Delta_c=0$.

In conclusion, we have illustrated the nature of coherent
spin dynamics driven by quantum-fluctuations in small condensates.
The time evolution of population of atoms at different hyperfine spin states
is shown to reveal intrinsic dynamics of quantum fluctuations of
order parameters and spin projections.
These dynamics can be probed by studying eigenfrequencies
of a coupled condensate-photon system in a
high-finesse optical cavity available in laboratories.
We thank Gerard Milburn, Junliang Song for stimulating discussions.
This work is in part supported by NSFC, $973$-Project (China), and NSERC
(Canada), CIFAR and A. P. Sloan foundation.

\end{document}